# Collective states of interacting ferromagnetic nanoparticles

O. Petracic,[a] X. Chen,[a] S. Bedanta,[a] W. Kleemann,[a,*] S. Sahoo,[b]
S. Cardoso,[c] and P. P. Freitas[c]

[a]*Angewandte Physik, Universität Duisburg-Essen, D-47048 Duisburg, Germany*

[b]*Department of Physics and Astronomy, University of Nebraska, Lincoln, Nebraska 68588, USA*

[c]*INESC, Rua Alves Redol 9-1, 1000 Lisbon, Portugal*

**Abstract**

Discontinuous magnetic multilayers [CoFe/Al$_2$O$_3$] are studied by use of magnetometry, susceptometry and numeric simulations. Soft ferromagnetic Co$_{80}$Fe$_{20}$ nanoparticles are embedded in a diamagnetic insulating a-Al$_2$O$_3$ matrix and can be considered as homogeneously magnetized superspins exhibiting randomness of size (*viz.* moment), position and anisotropy. Lacking intra-particle core-surface ordering, generic freezing processes into collective states rather than individual particle blocking are encountered. With increasing particle density one observes first superspin glass and then superferromagnetic domain state behavior. The phase diagram resembles that of a dilute disordered ferromagnet. Criteria for the identification of the individual phases are given.

*PACS*: 75.10.Nr; 75.50.Lk; 75.40.Gb; 75.60.Ej

*Keywords:* Magnetic nanoparticles; Dipolar interaction; Superspin glass; Superferromagnetism

## 1. Introduction

The physics of nanoscale magnetic materials is a vivid subject in current magnetism research. This is partially due to the promised potential in modern data storage applications [1] but mainly due to the wide spectrum of novel effects found [2]. One particularly interesting topic is the study of assemblies of magnetic nanoparticles, where each particle is in a magnetic single-domain state. In the simplest case of ferromagnetic (FM) particles showing coherent reversal of the moments, one can assign to each granule a single moment or "superspin" being usually in the order of 1000 $\mu_B$, where $\mu_B$ is Bohr's magneton. A FM nanoparticle is defined as superparamagnetic (SPM), when the energy barrier, $E_B$, for a magnetization reversal is comparable to the thermal energy, $k_BT$, during the measurement. The direction of the superspin then fluctuates with a frequency $f$ or a characteristic relaxation time, $\tau^{-1} = 2\pi f$. It is given by the Néel-Brown expression [3],

$$\tau = \tau_0 \exp(KV / k_BT), \qquad (1)$$

where $\tau_0 \sim 10^{-10}$ s is the inverse attempt frequency, $K$ an effective anisotropy constant and $V$ the volume of the nanoparticle. The energy barrier is here approximated by $E_B = KV$.

The magnetic behavior of the particle is characterized by the so-called "blocking" temperature, $T_b$, below which the particle moments appear blocked in the time scale of the measurement, $\tau_m$. This is the case, when $\tau_m \approx \tau$. Using Eq. (1) one obtains

$$T_b \approx KV / k_B \ln(\tau_m/\tau_0). \qquad (2)$$

An ensemble of nanoparticles is denoted as SPM, when the magnetic interactions between the particles are sufficiently

---

small [4]. Then the magnetic behavior of the ensemble is essentially given by the configurational average over a set of independent particles. More generally, one can denote the magnetic behavior as SPM in the sense of a thermodynamic phase. No collective inter-particle order exists, while the intra-particle spin structure is FM ordered. In the case of small concentrations of particles, only SPM behavior is observed.

However, for increasing concentrations the role of magnetic interactions becomes non-negligible. The mean (point) dipolar energy of two interacting nanoparticles, *e.g.* each with a moment of $\mu = 3000\mu_B$ and center-to-center distance of $D = 6$ nm yields $E_{d-d}/k_B = (\mu_0/4\pi k_B) \mu^2 / D^3 = 26$ K. Considering all neighbors, it is obvious that the effect of dipolar interactions can be observed even at temperatures in the order of 100 K. In the case of imperfectly spherical particles one also needs to take into account higher-order multipole terms [5,6]. Consequently, with increasing particle density one finds a crossover from single-particle blocking to collective freezing [2,4]. One can distinguish two kinds of collective states. For an intermediate strength of dipolar interactions, randomness of particle positions and sufficiently narrow size distribution one can observe a superspin glass (SSG) state. Here the superspins freeze collectively into a spin glass phase below a critical temperature, $T_g$ [2,7–9]. For higher densities of particles and hence stronger interactions, one can observe a superferromagnetic (SFM) state. It is characterized by a FM inter-particle correlation [8, 10–12].

There exist various experimental realizations of magnetic nanoparticle assemblies, *e.g.* frozen ferrofluids [2], discontinuous metal-insulator multilayers (DMIMs) [8,13], co-sputtered metal-insulator films [14], self-organized particle arrays on surfaces [15], focused ion-beam structured thin films [16] and mechanically alloyed materials [17].

In this article we discuss experimental studies on a series of DMIMs, $[Co_{80}Fe_{20}(t_n)/Al_2O_3(3nm)]_{10}$, where the nominal thickness is varied in the range $0.5 \leq t_n \leq 1.8$ nm. A crossover from SSG to SFM behavior is observed and a phase diagram, *i.e.* transition temperature *vs.* nominal thickness, is constructed. Moreover, results from numeric simulations on a SFM system are presented.

**2. Details of experiment and simulation**

We performed experimental studies on DMIMs, $[Co_{80}Fe_{20}(t_n)/Al_2O_3(3nm)]_{10}$, with nominal thickness in the range $0.5 \leq t_n \leq 1.8$ nm. The samples were prepared by sequential Xe-ion beam sputtering from $Co_{80}Fe_{20}$ and $Al_2O_3$ targets on glass substrates [18]. Due to the non-wetting properties of the soft FM $Co_{80}Fe_{20}$, it grows as an ensemble of nearly spherical particles being eventually embedded in the insulating $Al_2O_3$ matrix. The average diameter of CoFe particles can be tuned by the nominal thickness, $t_n$. *E.g.* for $t_n = 0.9$ nm the mean diameter of CoFe particles is $\langle D \rangle = 2.8$ nm [19]. During growth a weak magnetic field ($\mu_0 H \sim 10$ mT) was applied in order to induce an in-plane easy-axis. Magnetization and *ac* susceptibility measurements were performed by use of a commercial superconducting quantum interference device (SQUID) magnetometer (Quantum Design, MPMS-5S).

Ensembles of dipolarly interacting nanoparticles were studied by Monte-Carlo simulations with the Metropolis algorithm [20]. The simulation essentially minimizes the total energy of the entire system at a given field, $H$, parallel to the $z$ axis and temperature, $T$. The energy of a particle $i$ is given by

$$E_i = -K V_i (\mathbf{k}_i \cdot \mathbf{s}_i)^2 - \mu_0 M_s V_i H s_i^z \qquad (3)$$
$$+ (\mu_0 M_s^2/4\pi) \sum_{\{j\}} V_i V_j [\mathbf{s}_i \cdot \mathbf{s}_j - 3(\mathbf{s}_i \cdot \boldsymbol{\rho}_{ij})(\mathbf{s}_j \cdot \boldsymbol{\rho}_{ij})] / r_{ij}^3,$$

where $K$ is the effective anisotropy constant, $V_i$ the volume and $M_s$ the saturation magnetization of a particle. $\mathbf{k}_i$, $\mathbf{s}_i$ and $\boldsymbol{\rho}_{ij}$ are the unit vectors of the anisotropy axis, the superspin direction and the distance vector to particle $j$, $\boldsymbol{\rho}_{ij} = \mathbf{r}_{ij} / |\mathbf{r}_{ij}|$, respectively. The summation in the third term includes all particles $\{j\}$ surrounding $i$ within a cut-off radius of 25 nm. In this study we focused on the spatial superspin configuration of a 2D array of nanoparticles after proper equilibration of the ensemble, *i.e.* 20000 Monte Carlo steps. The positions of approximately 100 particles were directly transferred from transmission electron microscopy (TEM) images of a SSG DMIM sample with $t_n = 0.9$ nm [19]. The volumes $V_i$ were calculated from the diameters as found from the TEM images assuming spherical particles. Values for the effective anisotropy were $K = 4 \cdot 10^5$ J/m$^3$ [19] and for the saturation magnetization $M_s = 1.44$ MA/m (bulk Co).

In order to simulate a dense SFM system ($t_n = 1.4$ nm), the particle volumes were manually scaled by a factor $1.5^3 \approx (1.4 \text{ nm}/0.9 \text{ nm})^3$ under the assumption of Vollmer-Weber particle growth with constant areal density of growth nuclei [8,21]. Moreover, an easy in-plane axis was introduced by choosing random anisotropy axis vectors from a 3D unit sphere, where the probability for the $x$-, $y$-, and $z$-component corresponds to a Gaussian function with width $\sigma_y = 0.6$ for the $y$ and $\sigma_z = 0.3$ for the $z$-component.

**3. Phase diagram of interacting nanoparticles**

A cartoon of the expected phase diagram, transition temperature *vs.* nominal thickness, is depicted in Fig. 1. At high temperatures, $T > T_{c,bulk}$ (*i.e.* bulk Curie temperature) the system is paramagnetic (PM). Below $T_{c,bulk}$ spontaneous FM order builds up in each particle. Here one should note that a possible finite size effect of the Curie temperature might be included. At very small nominal thicknesses and, hence, small concentrations, the system behaves SPM. Any possible effect of inter-particle interactions is not apparent, since the single-particle blocking at $T_b = T_b(t_n, \tau_m)$ disguises any transition at temperatures $T < T_b$.

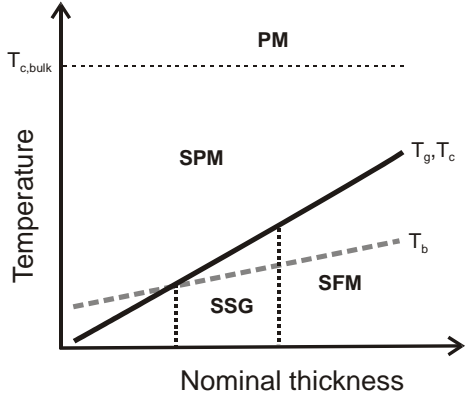

FIG. 1. Schematic phase diagram, transition temperature *vs.* nominal thickness, with paramagnetic (PM), superparamagnetic (SPM), superspin glass (SSG) and superferromagnetic (SFM) phase. Relevant lines are the blocking temperature of the individual particles, $T_b$, and the collective transition line, *i.e.* the glass transition, $T_g$, or SFM transition temperature, $T_c$.

*3.1. Superspin glass (SSG) ordering*

For higher nominal thicknesses collective inter-particle ordering occurs, where the ordering temperature exceeds the blocking temperature. For particle arrangements exhibiting randomness of size (*viz.* moment), position and anisotropy first a SSG phase is encountered. The transition line, $T_g = T_g(t_n)$, is a phase transition line separating the SPM and SSG phase.

Evidence for SSG ordering in nanoparticle assemblies is found in several systems [2,7–9] using a wide spectrum of various methods, *i.e.*

(1) A first simple test to identify SSG ordering is achieved from zero-field-cooled / field-cooled (ZFC/FC) curves of the magnetization, $M$ vs. $T$. Fig. 2 shows $M_{ZFC}$ and $M_{FC}$ *vs.* $T$ obtained on the SSG system [Co$_{80}$Fe$_{20}$($t_n$)/Al$_2$O$_3$(3nm)]$_{10}$ with $t_n = 0.9$ nm. While a typical peak in the ZFC curve is found in both SPM and SSG samples, the observation of a *decrease* in $M_{FC}(T)$ upon cooling (see arrow) can only be observed in SSG systems [22]. The appearance of a dip in this example is due to an additional paramagnetic signal from small clusters or atoms dispersed between the particles [21].

(2) SSG and more generally any spin glass system exhibits the so-called aging effect in the ZFC magnetization [23]. Here the sample is rapidly cooled down in zero field to a temperature below $T_g$ and a certain waiting time, $t_w$, spent at this temperature. Finally, a small field is applied and the magnetization recorded as function of time, $t$. Then the derivative $S(t) = (1/\mu_0 H)(\partial M(t)/\partial \ln t)$ shows a peak at $t \approx t_w$ [2,9,22]. Hence, the system "remembers" the time spent in zero field before the field is applied.

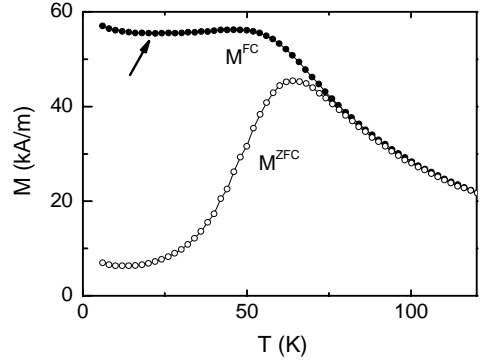

FIG. 2. ZFC magnetization, $M_{ZFC}$, and FC magnetization, $M_{FC}$, *vs.* $T$ of [Co$_{80}$Fe$_{20}$(0.9nm)/Al$_2$O$_3$(3nm)]$_{10}$ measured in a field of 0.4 mT. The arrow marks a dip in $M_{FC}$ being typical of SSG systems.

(3) A similar experiment reveals the so-called memory effect [23]. The sample is rapidly cooled in zero field to a certain stop temperature, $T_s < T_g$. After waiting a certain time at $T_s$ the cooling procedure is resumed to some lower temperature. Then a small field is applied and the ZFC magnetization curve, $M_{ZFC}(T)$, recorded upon heating up. The obtained ZFC curve shows a dip exactly at $T_s$. This is best seen in the difference between the curves without and with intermittent stop, $\Delta M = M_{ZFC}^{ref} - M_{ZFC}$ [2,9,22]. Fig. 3 shows results on the SSG system with $t_n = 0.9$ nm, where a stop of $2 \cdot 10^4$ s at $T_s = 32$ K was applied.

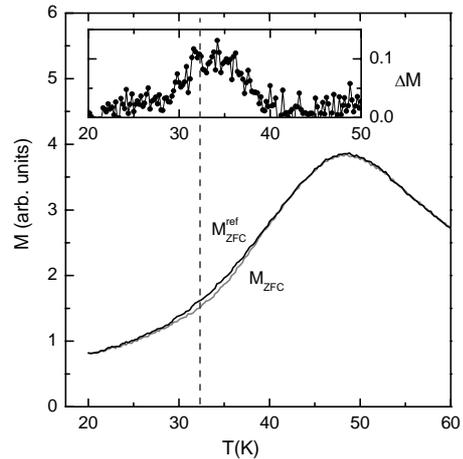

FIG. 3. Memory effect observed in $M_{ZFC}(T)$ in a field of 0.01 mT on the SSG sample [Co$_{80}$Fe$_{20}$(0.9nm)/Al$_2$O$_3$(3nm)]$_{10}$. A stop of duration $2 \cdot 10^4$ s was applied at 32 K. The inset shows the difference curve, $\Delta M = M_{ZFC}^{ref} - M_{ZFC}$.

(4) Spin glasses exhibit a divergence of the non-linear susceptibility, $\chi_3$, at the critical temperature, $T_g$, i.e. [23]

$$\chi_3 \propto (T/T_g - 1)^{-\gamma}, \qquad (4)$$

where $\gamma$ is the corresponding critical exponent. $\chi_3$ is obtained from isothermal magnetization curves, which can be expanded as odd power series, $M = \chi_1 H - \chi_3 H^3 + O(H^5)$. From fits of the $\chi_3(T)$ values to Eq. (4) one can then extract $T_g$ and $\gamma$. SSG systems typically exhibit glass temperatures in the range $T_g \sim 50$ K [8,24].

(5) Another test for SSG ordering is based on measurements of the complex ac susceptibility, $\chi = \chi'-i\chi''$, in zero field. The real part, $\chi'(T)$ shows a peak at the "freezing" temperature $T_f$. With decreasing ac frequency, $f$, the peak position, $T_f$, shifts to lower temperatures. In a SSG there will be a limit value, $T_f \to T_g$ for $f \to 0$. This is expressed as a critical power law [23,25],

$$\tau = \tau^* (T_f/T_g - 1)^{-z\nu}, \qquad (5)$$

where $\tau$ is the characteristic relaxation time of the system being probed by the ac frequency, $2\pi f = \tau^{-1}$, $\tau^*$ is the relaxation time of the individual particle moment, $\nu$ the critical exponent of the correlation length, $\xi = (T/T_g-1)^{-\nu}$, and $z$ relates $\tau$ and $\xi$ as $\tau \propto \xi^z$. E.g. in the DMIM with $t_n = 0.9$ nm we found reasonable values, $T_g = 61\pm 2$ K, $z\nu = 10.2$ and $\tau^* = 10^{-8 \pm 1}$ s [19].

(6) One can achieve data collapse of the ac susceptibility on one master curve in the dynamical scaling plot, $(T/T_g - 1)^{-\beta}\chi''/\chi_{eq}$ vs. $\omega\tau^*(T/T_g - 1)^{-z\nu}$, and thus obtain values for $T_g$, $z\nu$, and $\beta$. Here $\chi_{eq}(T)$ is the hypothetical equilibrium curve, which can be approximated by a Curie-Weiss hyperbola at high temperatures, $\beta$ the critical exponent of the order parameter and $\omega = 2\pi f$ [24].

(7) In the case of a SSG system, the so-called Cole-Cole plot, $\chi''$ vs. $\chi'$, shows a strongly flattened (shifted) semi-circle [23]. An example is presented in Fig. 4 (a) showing results on a DMIM with $t_n = 0.9$ nm [26]. In contrast, a SPM system would exhibit an almost complete (not shifted) semi-circle [27].

### 3.2. Superferromagnetic (SFM) ordering

If the nominal thickness or the particle concentration is increased, higher-order multipole terms of the dipolar interaction become relevant [5,6]. This can lead to SFM ordering [10,11,28] (see Fig. 1), where regions of FM correlated particle moments (SFM domains) are found [12,29]. The magnetic behavior is then essentially given by the nucleation and motion of domain walls (DWs). We assume such a SFM domain state to occur in our DMIMs for $t_n > 1.1$ nm. Analogously to SSG systems, there exist pertinent methods to identify SFM behavior in nanoparticle systems.

(1) The Cole-Cole plot, $\chi''$ vs. $\chi'$, appears to be completely dissimilar to that of a SPM and SSG system [27]. An experimental Cole-Cole plot of a SFM DMIM with $t_n = 1.3$ nm is shown in Fig. 4 (b). For low frequencies (right hand side) one finds a quarter circle, whereas at high $f$ an increasing part with positive curvature appears. These features can be successfully modeled in the framework of the motion of a pinned DW in a random FM exhibiting *relaxation*, *creep*, *slide* and *switching* [11,27,28,30].

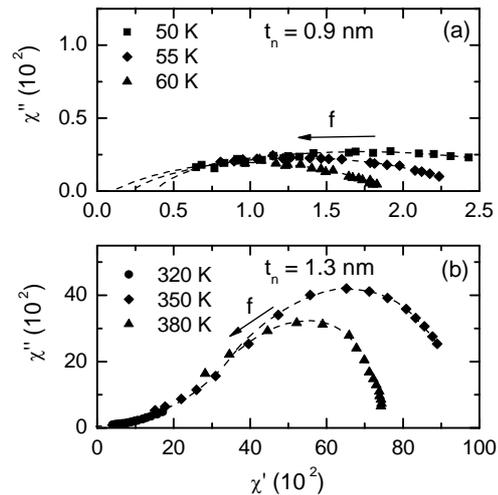

FIG. 4. Cole-Cole plots, $\chi''(f)$ vs. $\chi'(f)$, for a SSG DMIM sample with $t_n = 0.9$ nm (a) at $T = 50$, 55 and 60 K and for a SFM DMIM with $t_n = 1.3$ nm (b) at $T = 320$, 350 and 380 K. The arrows indicate the order of ac frequencies corresponding to each data point. Note that both plots are deliberately in a 1:1 scale for the $\chi''$ and $\chi'$ axes.

(2) Another strong indication of SFM behavior can be found from analyzing relaxation curves of the thermo-remanent magnetic moment, $m$ vs. $t$. Ulrich et al.[31] introduce a universal relaxation behavior for any nanoparticle system according to a decay law of the form

$$(d/dt)\, m(t) = -W(t)\cdot m(t). \qquad (6)$$

Here the time dependent logarithmic decay rate, $W(t) = -(d/dt) \ln m(t)$, obeys a power law,

$$W(t) = At^{-n}, \text{ with } t \geq t_0. \qquad (7)$$

A SFM system can be distinguished from a SSG by the exponent $n$. $A$ is an arbitrary constant. While SSG samples exhibit $n < 1$, SFM systems appear to have values $n > 1$ [32]. In the latter case $m(t)$ obeys a power law,

$$m(t) = m_\infty + m_1 t^{1-n}, \qquad (8)$$

where $m_\infty$ and $m_1$ are proper fit constants. We find excellent agreement with extracted values for $n$ from experimental relaxation curves in our DMIM samples [32].

(3) Relaxation curves, $m(t)$, of a SFM sometimes show a peculiar behavior after rapidly FC the sample. One can

observe an intermediate *increase* of $m(t)$ after switching *off* the field [33]. In this case the relaxation is characterized by two processes. First, a decay according to a power law being due to DW motion. Second, an increasing contribution according to a stretched exponential due to the post-alignment of particle moments *inside* the SFM domains, *i.e.*

$$m(t) = m_0 + m_1 t^{1-n} + m_2[1-\exp(-(t/\tau)^\beta)], \quad (9)$$

where $m_0$, $m_1$ and $m_2$ are fit constants. $\tau$ and $\beta$ are the effective relaxation time and exponent, respectively, of the post-alignment process.

Similar to regular FM, the transition temperature, $T_c$, can be obtained from analyzing hysteresis loops, *i.e.* recording the remanent and coercive fields *vs.* $T$. Finally, the results on the DMIMs can be summarized in an experimental phase diagram as presented in Fig. 5 resembling that of a dilute disordered ferromagnet [34]. One finds SSG behavior for a nominal thickness $t_n <$ 1.1 nm, while a SFM state is encountered at $t_n >$ 1.1 nm. Moreover, a crossover to a random field domain state (RFDS) is found at lower temperatures [32] in the SFM state. The borders of this phase diagram have recently been specified to be $0.5 < t_n < 0.7$ nm for SPM-to-SSG [21] and $1.6 < t_n < 1.8$ nm for SFM-to-percolated FM [35].

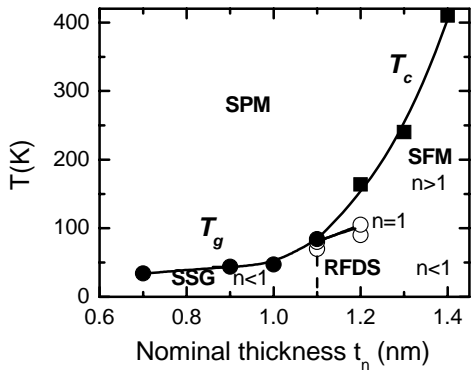

FIG. 5. Experimental phase diagram of DMIMs with nominal thickness $t_n$ and phases SPM, SSG, SFM and RFDS. Values for the exponent $n$ from the analysis of relaxation curves $m(t)$ are shown (see text).

In addition, Monte-Carlo simulations on a 2D array of dipolar interacting nanoparticles were performed in order to study SFM ordering. Fig. 6 shows the spin structure of approximately 100 particles in zero field after ZFC from 300 to 5 K. The simulation area is $50 \times 50$ nm$^2$. One observes FM correlated regions including several particles [12,29], which can be interpreted as SFM domains. This study neglects higher-order multipole terms of the dipolar interaction [5,6]. Probably a more advanced simulation including those contributions would drastically enhance the SFM ordering tendency.

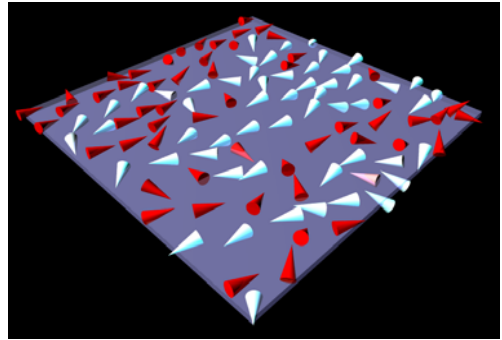

FIG. 6. Spin structure at zero field of a 2D array of dipolar interacting nanoparticles after ZFC from 300 to 5 K. The particle moments are represented as cones. The color code (red, white) corresponds to the *x*-component of the moment.

## 4. Conclusion

Ensembles of ferromagnetic nanoparticles can exhibit collective behavior. When the dipolar interaction becomes non-negligible, superparamagnetism is overcome below a certain threshold inter-particle clearance (at constant density). In our granular discontinuous metal-insulator multilayers, $[Co_{80}Fe_{20}(t_n)/Al_2O_3(3nm)]_{10}$ we can observe supersnpin glass and superferromagnetic behavior at $t_n >$ 0.5 nm and $>$ 1.1 nm, respectively. Both collective states are clearly identified by various kinds of experimental methods.

## Acknowledgment

Thanks are due to the DFG (KL306/37-1, and Graduate School GK 277) for financial support.